\documentclass[11pt]{article}
\usepackage{amssymb, latexsym, amsmath, amsthm, amsfonts, verbatim, bm}
\usepackage[final]{graphicx}
\usepackage{xcolor}
\usepackage{rotating}
\usepackage{pstricks}
\usepackage{pst-plot}
\usepackage{pst-all}

\begin{document}


\newcommand{\B}[1]{\textbf{#1}}
\newcommand{\ra}{\rightarrow}
\newcommand{\sinc}{\text{sinc}}
\newcommand{\supp}{\text{supp}}
\newcommand{\suppe}{\text{supp}_{\varepsilon}}
\newcommand{\ind}{\B{1}}
\newcommand{\sgn}{\text{sgn}}
\newcommand{\etal}{\text{et al. }}
\newcommand{\Beta}{{\boldsymbol\beta}}
\newcommand{\Radon}{\mathcal{R}}
\newcommand{\Xray}{\mathcal{P}}
\newcommand{\Fourier}{\mathcal{F}}
\newcommand{\Hilbert}{\mathcal{H}}
\newcommand{\Identity}{\mathcal{I}}
\newcommand{\BL}{\mathcal{L}}
\newcommand{\R}{\mathcal{R}}
\newcommand{\Cov}{\text{Cov}}
\newcommand{\argmin}{\operatorname{argmin}}

\title{Methods for Few-View CT Image Reconstruction}
\author{Kyle M. Champley, Michael B. Zellner, \\Joseph W. Tringe, and Harry E. Martz Jr.}
\maketitle

\begin{abstract}
Computed Tomography (CT) is an essential non-destructive three dimensional imaging modality used in medicine, security screening, and inspection of manufactured components.  Typical CT data acquisition entails the collection of a thousand or more projections through the object under investigation through a range of angles covering one hundred eighty degrees or more.  It may be desirable or required that the number of projections angles be reduced by one or two orders of magnitude for reasons such as acquisition time or dose.  Unless specialized reconstruction algorithms are applied, reconstructing with fewer views will result in streak artifacts and failure to resolve object boundaries at certain orientations.  These artifacts may substantially diminish the usefulness of the reconstructed CT volumes.

Here we develop constrained and regularized numerical optimization methods to reconstruct CT volumes from 4-28 projections.  These methods entail utilization of novel data fidelity and convex and non-convex regularization terms.  In addition, the methods outlined here are usually carried out by a sequence of two or three numerical optimization methods in sequence.

The efficacy of our methods is demonstrated on four measured and three simulated few-view CT data sets.  We show that these methods outperform other state of the art few-view numerical optimization methods.
\end{abstract}

\section{Introduction and Background}

X-ray Computed Tomography (CT) provides a method to non-destructively produce cross-sectional images of the linear attenuation coefficients (units of inverse length) of materials through a specific volume of interest \cite{Martz_XrayBook_2017}.  This is accomplished by measuring the attenuation of x-rays through the object with a collection of one or more linear or panel detectors.  Then CT image reconstruction algorithms are applied to the measured data to form the linear attenuation coefficient images.

The collection of rays from a source to the x-ray detector is referred to as a projection or view.  Multiple projections at different angles must be measured to properly reconstruct an object.  Different trajectories are measured by rotating the object, rotating the source and detectors, or by using multiple stationary sources and detectors of varying perspective.  Conventional CT imaging requires that the number of projections be roughly equal to the number of detector columns in a projection.  When the number of projections is one or more orders of magnitude smaller than this, streak artifacts pollute the reconstructed image, unless specialized algorithms are applied to suppress these artifacts.

The number of projections may be limited for many reasons.  When scanning an object in motion, one may deliberately collect a limited number of views to shorten the data acquisition time to reduce the artifacts associated with object motion blur.  Some CT protocols require that the dose delivered to the object or patient be limited.  Reducing the number of measured projections is one way to reduce dose.  Flux-limited systems employ long integration times per projection to measure adequate photon statistics.  The flux may be limited by dim sources (low current), large source-to-detector distances, small detector pixels, or highly attenuating objects.  A reduced number of projections may be desired in these systems to increase throughput.  One final scenario where few-view CT is common is the space of fixed gantry systems.  These systems employ multiple source-detector pairs mounted on a stationary gantry.  These systems are commonly used in airport security, especially those scanners designed to detect explosives and other contraband in carry-on luggage or in flash x-ray systems \cite{Zellner_MEFCT_2018} which are designed to image highly dynamic events such as explosions or ballistic impacts.

In this paper we discuss a few strategies for reconstructing few-view CT data.  Some of the methods developed in this paper are applicable to only a certain class of objects because they take advantage of specific assumptions for the image properties one expects.

We will not discuss Deep Learning (DL) method for few-view CT reconstruction in this paper which have found great interest in the past several years.



\section{Notation and Conventions}

In the following we shall denote the measured attenuation data by $g$ and the volume of linear attenuation coefficients by $f$.  The linear forward model, known as the X-ray Transform, is given by $P$.  Thus the goal of CT image reconstruction is to find an $f$ such that $Pf \approx g$.  The adjoint of the X-ray Transform operator, given by $P^T$, is referred to as the backprojection operator.

\section{Challenges of Few-View CT Image Reconstruction \\ and Proposed Solutions}

Mathematically, few-view CT reconstruction is a problem of solving an underdetermined system of linear equations in the presence of noisy and corrupted (e.g. beam hardening, scatter, etc.) measurements.  Underdetermined systems have a non-empty null space.  A function, $n$, is in the null space of $P$ if and only if $Pn = 0$.  A non-empty null space implies that there are multiple solutions to a given set of equations.  For example, suppose that $\{n_k\}_{k=1}^M$ is a set of functions in the null space of $P$ and suppose that $f$ is a function such that $Pf = g$.  Then, by definition, $f + \sum_{k=1}^M a_k n_k$, where $a_k \in \mathbb{R}$ are also solutions.  Thus there are an infinite number of solutions to $Pf = g$ and unless one makes some assumptions regarding the types of functions that can be considered as solutions, one has no way of determining which solution is the true solution.  The leading strategy for few-view CT reconstruction includes constraining the solution space of the inverse problem to reduce the dimensionality of the null space of the forward operator, $P$.  These reconstruction algorithms are iterative: they generate a series of possible solutions where each iteration brings one closer to the desired solution.  The associated algorithms are usually taken from either the field of numerical linear algebra or the field of numerical optimization.

The reconstruction solution is a map of linear attenuation coefficient (LAC) values, which are, by definition, nonnegative.  Thus one common strategy is to constrain the solution to be nonnegative.

Real objects have structure and most regions are quasi-homogeneous.  Thus another common method includes constraining the solution to meet some smoothness criteria.  Many of these methods measure image roughness by a loss function applied to the magnitude of the gradient of the volume.  Reducing image roughness is mathematically similar to image denoising.  Thus we may leverage the algorithms from the rich subject of image denoising.  Moreover, it has been known for some time \cite{Cormack_ApplPhys_1963,Logan_DukeMath_1975,Louis_MathZ_1984,Louis_SIAM_1984} that highly oscillatory functions occupy a large portion of the null space of the X-ray Transform, even when acquiring large numbers of projections.  These highly oscillatory functions often contain streak artifacts which coincide with measured paths of the X-ray Transform.  Thus it is not surprising that image denoising methods have been so effective at reducing artifacts in CT images.  Two  algorithms that we have found particularly effective are the Adaptive Steepest Descent- Projection onto Convex Sets (ASD-POCS) \cite{Pan_PhysMedBiol_2008} and the Improved Total Variation Constrained Reconstruction (iTV) \cite{Kachelriess_PhysMedBiol_2011} methods.  The iTV method can be seen as a modification of ASD-POCS.  These methods alternate between iterations of Simultaneous Algebraic Reconstruction Technique (SART) \cite{Kak_UltrasonImag_1984} and iterations that reduce the Total Variation (TV) norm \cite{ROF_TV_1992} in a clever way.  An alternative to SART, called the Multiplicative Algebraic Reconstruction Technique (MART) \cite{Herman_TheorBiol_1970,Gordon_PhysMedBiol_2004} has also been shown to be effective in few-view CT image reconstruction \cite{Moser_MeasSciTechnol_2014}.

Another class of few view reconstruction methods entails reducing the number of unknowns.  In most cases, enlargement of the reconstructed image voxel size is not useful because it leads to unacceptably poor image resolution, and creates images that do not adequately describe the data.  Many image features such as boundaries between materials cannot be described by low resolution images.  One strategy is to model the image by an adaptable mesh \cite{Gullberg_TMI_2006}, where the number of nodes used varies with the estimated local variation.  Another strategy entails the use of a so-called image patch dictionary \cite{Aharon_TIP_2006}.  In this method, one constructs a library of image patches and estimates the CT image by building an image where each patch in the image is composed of a linear combination of a small number of image patches in the dictionary \cite{Wang_TMI_2012}.

It is also possible to leverage a prior image such as an earlier scan of a patient or object with the Prior Image Constrained Compressed Sensing (PICCS) algorithm \cite{Chen_MedPhys_2008}. Compressed Sensing (CS) \cite{Donoho_TIT_2006} has been broadly popular for almost two decades.  Dictionary learning and TV denoising can also be classified as CS techniques.

It is difficult to develop mathematical formulas and metrics that quantify undesirable image features.  This remains one of the main challenges in developing new methods for few-view CT image reconstruction.

It should not be overlooked that it is not only possible to constraint the solution space, but also to develop algorithms that avoid introducing null functions in the iterative process.  In many of these iterative reconstruction algorithms, once a null function is introduced in the iterative process, it is difficult and perhaps even impossible to remove it.  To demonstrate this fact, consider solving the reconstruction by minimizing a least squares norm with a gradient-based method.  The least squares functional is given by $$\Phi(f) := \frac{1}{2}\| Pf-g\|_2^2.$$  The gradient of this cost functional is $$\Phi(f) = P^T(Pf-g).$$  Then a gradient-based optimization method takes the form $$f_{k+1} = f_k - Q_k(\Phi'(f_k)),$$ where $Q$ is some convex operation and $Q(0) = 0$.  Now suppose that $f_{LS}$ is the least solution and for some $k$, $f_k = f_{LS} + n$, where $n$ is a null function.  Then we have that
\begin{eqnarray*}
f_{k+1} &=& f_k - Q_f(\Phi'(f_k)) \\
&=& f_{LS} + n - Q_f(P^T(P(f_{LS}+n) - g)) \\
&=& f_{LS} + n - Q_n(P^T(Pf_{LS} - g)) \\
&=& f_{LS} + n - Q_n(0) \\
&=& f_{LS} + n
\end{eqnarray*}
and thus we see that for all subsequent iterations, the solution with continue to be corrupted by a null function.

We hypothesize that this illustrates one of the strengths of the ASD-POCS/ iTV approaches to few-view reconstruction.  In these algorithms, each SART step is followed by several (5-30) TV minimization steps. These multiple TV steps help to suppress null functions (streak artifacts) while the SART algorithm attempts to perform the reconstruction and thus these null functions have less of a chance to enter the iterative process.


\section{Regularized Weighted Least Squares}

A common method for CT image reconstruction entails the utilization of the Regularized Weighted Least Squares (RWLS) functional.  The solution to this minimization problem is given by
\begin{eqnarray*}
\widehat{f}_{RWLS} := \underset{f}{\operatorname{argmin}} \; \Phi_{RWLS}(f),
\end{eqnarray*}
where
\begin{eqnarray*}
\Phi_{RWLS}(f) &:=& \frac{1}{2}\left(Pf - g\right)^TW\left(Pf - g\right)+ \beta R(f), \\
\end{eqnarray*}
$R(\cdot)$ is a functional that penalizes non-smooth solutions, and $\beta \geq 0$ controls the strength of the regularization.  A common choice for this regularizer is
\begin{eqnarray*}
R_p(f) := \| h(\nabla f) \|,
\end{eqnarray*}
where $\nabla$ is the gradient operator and $h$ is some convex loss function.  The special cases $h(t) = |t|$ and $h(t) = t^2$ result in total variation (TV) and Tikhonov regularization, respectively.  The $W$ matrix may be an estimate for the variance of the noise in the measured data or some measure of the relative importance of each measurement (e.g., for purposes of metal artifact reduction).  By choosing $W = I$, where $I$ is the identity matrix, this reduces down to Regularized Least Squares (RLS).  The gradient and Hessian of this cost functional are given by
\begin{eqnarray*}
\Phi_{RWLS}'(f) &=& P^T W(Pf-g) + \beta R'(f) \\
\Phi_{RWLS}''(f) &=& P^TWP + \beta R''(f).
\end{eqnarray*}

\section{Novel Methods for Few-View CT Reconstruction}

In this section we develop novel methods for few-view CT reconstruction.  These methods can be used as either stand-alone methods or integrated into a multi-step process.  An outline of a three-step process follows.  In the first step of this three-step process, the CT volume is reconstructed with a standard few-view CT reconstruction method.  This initial reconstruction is used to seed the initial state in the second step, which reconstructs a preliminary image using some strong constraints/regularization.  These heavy constraints are imposed to suppress artifacts, but may simultaneously produce a bias on the result such as the loss of low contrast features.  This is employed as a second step because sometimes the regularization methods used are non-convex; thus it helps to start with a reconstruction that is close to the desired results.  In the final step, one performs a reconstruction with any iterative reconstruction algorithm, starting with the image reconstructed in the previous step.  In this final step one usually imposes light constraints to reduce bias.

The motivation for this three-step approach derives from the likelihood of the second stage reconstruction to be biased or incomplete because of its reliance on inaccurate assumptions.  Regardless, the difference between the measured data and the forward projection of the second stage reconstruction is designed to have significantly lower contrast and reduced high frequency content as compared with the original measured data.  By seeding the third stage reconstruction with this preliminary reconstruction one reduces the chance of introducing streaks and other artifacts, i.e., null functions, into the final reconstruction.  Moreover, since the final stage reconstruction is not being constrained by any of the assumptions that went into the preliminary stage reconstruction, one reduces the risk of biasing the final result with the strong assumptions that created the image in the second stage.

\subsection{Motivation from Fan-Beam CT Sampling Theory} \label{sec:samplingTheory}

\begin{figure}[h!]
\begin{center}
\begin{tabular}{ccc}
\includegraphics[width=0.3\textwidth]{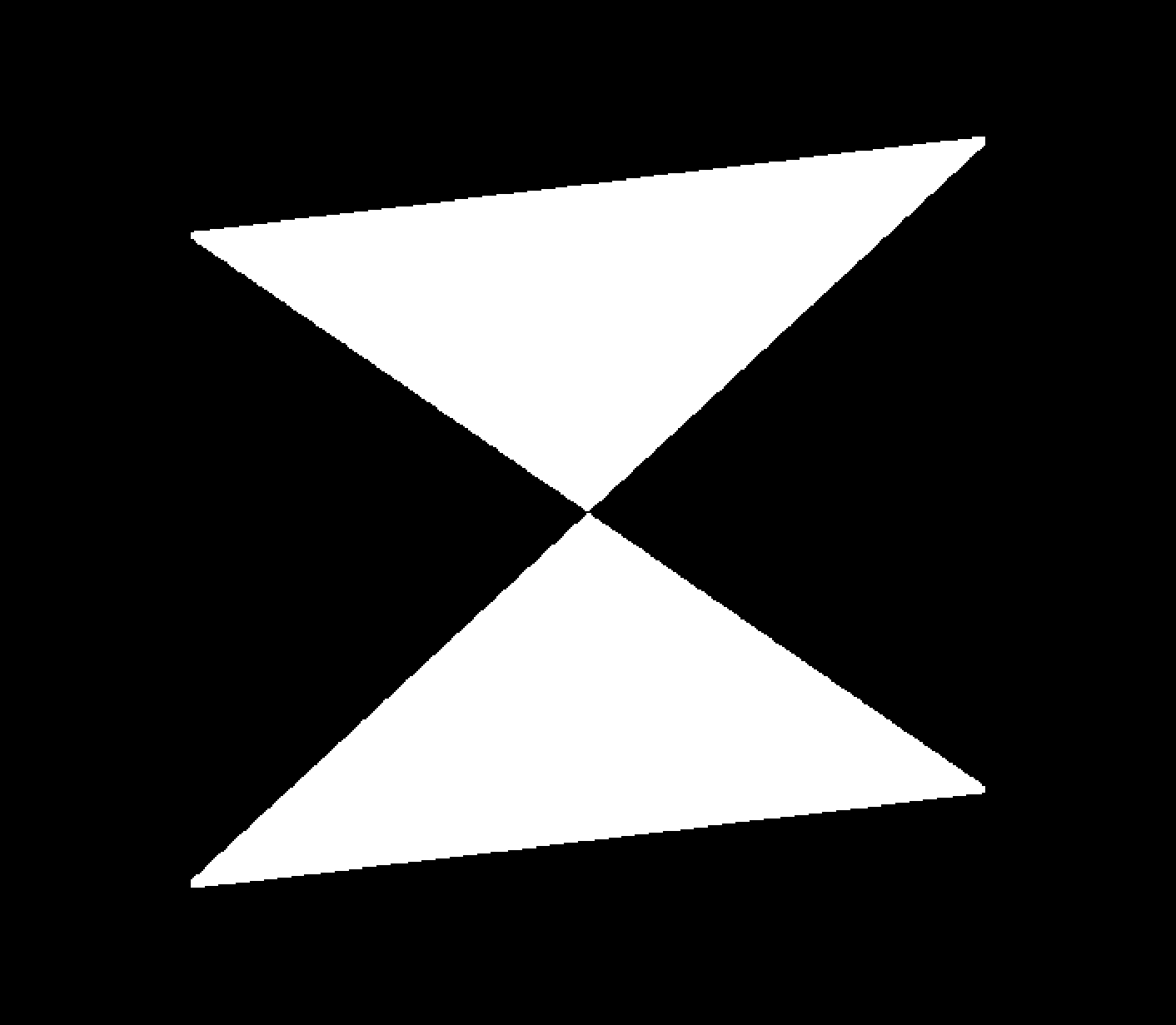}
& \includegraphics[width=0.3\textwidth]{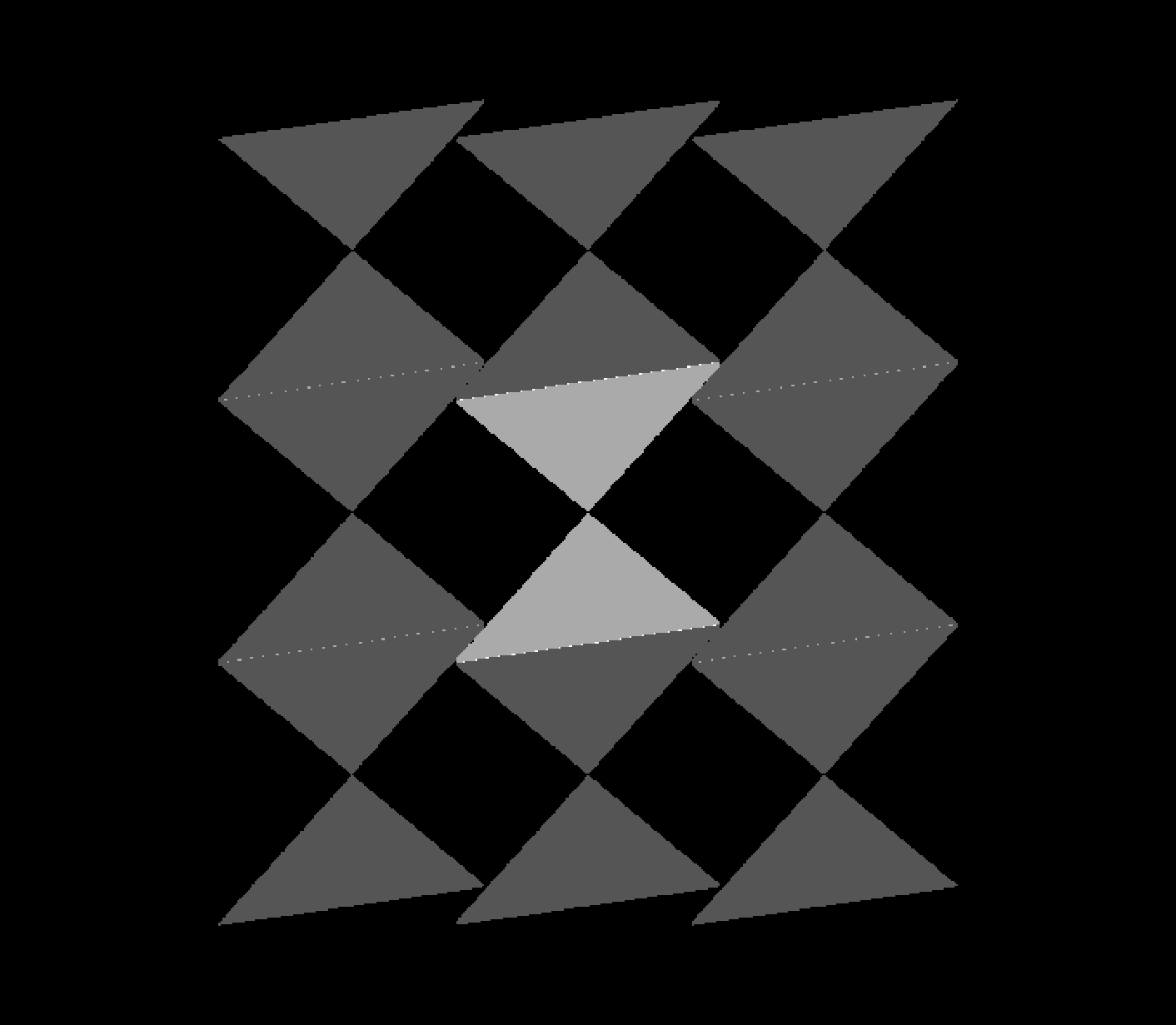}
& \includegraphics[width=0.3\textwidth]{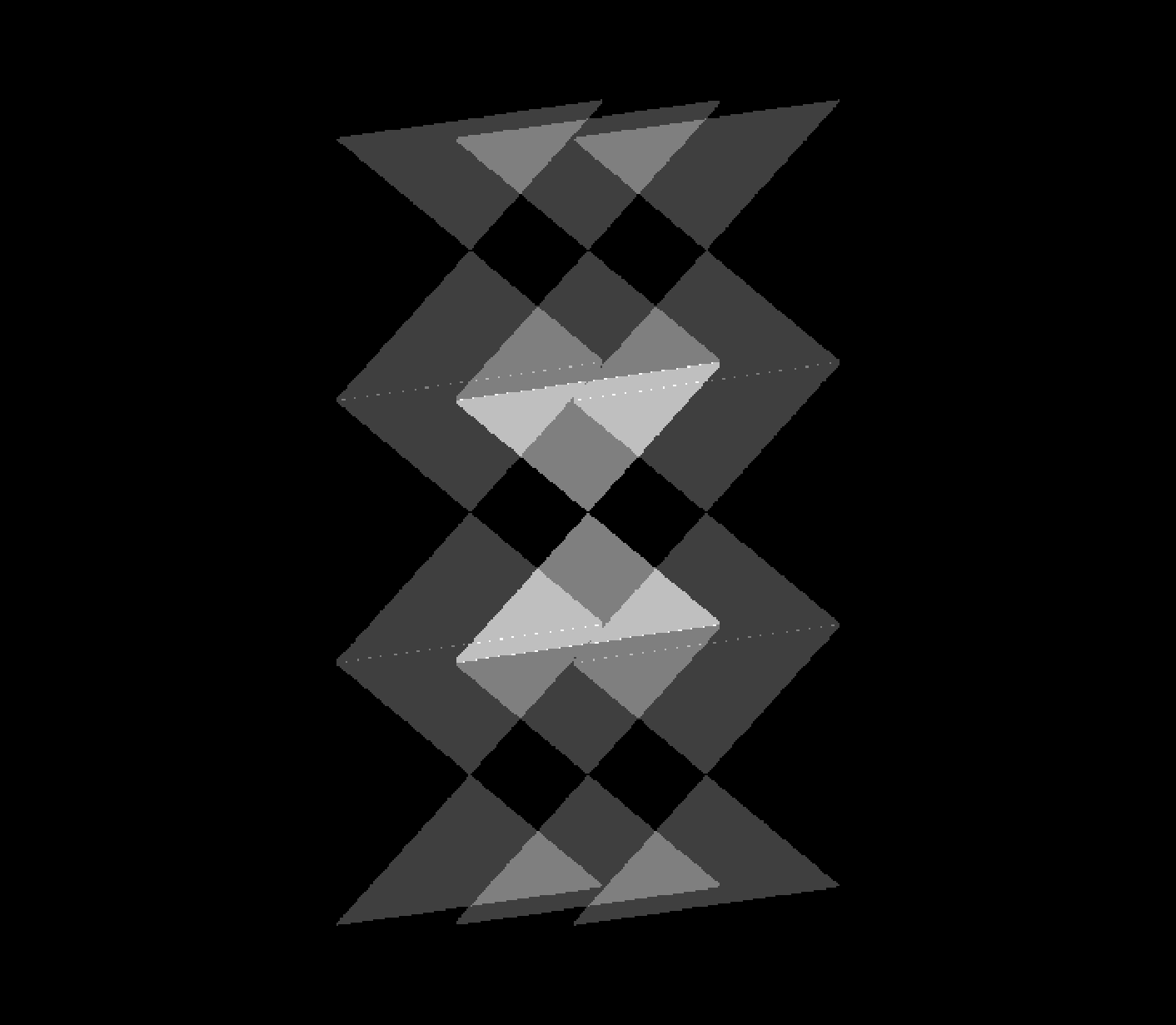}
\end{tabular}
\end{center}
\caption{Sampling the fan-beam X-ray Transform.  (left) Essential support of the 2D Fourier Transform of the fan-beam X-ray Transform.  The vertical direction is the frequency variable for $u$ (detector column) and the horizontal direction is the frequency variable for $\beta$ (projection angle).  Aliased copies of the essential support of the 2D Fourier Transform of the fan-beam X-ray Transform for the cases when the angles are (center) sufficiently sampled and (right) under-sampled.} \label{fig:fanBeamSampling}
\end{figure}

Our methods are partially motivated by CT sampling theory, which now briefly outline.  Consider the X-ray Transform in flat panel fan-beam coordinates
\begin{eqnarray*}
P_{FB}f(u, \beta) &:=& \int_\mathbb{R} f \left( R\theta(\beta) - \frac{l}{\sqrt{1+u^2}}\left[ \theta(\beta) - u\theta^\perp(\beta) \right] \right) \, dl,
\end{eqnarray*}
where $\theta(\beta) = (\cos\beta, \sin\beta)^T$ and $\theta^\perp(\beta) = \theta\left(\beta+\frac{\pi}{2}\right)$. It has been shown that the essential support of the 2D Fourier Transform of the fan-beam X-ray Transform \cite{Faridani_AM_2006, Izen_SIAM_2012} is supported in a bowtie-shaped region as shown in the left image of Figure \ref{fig:fanBeamSampling}.  The aperture of this bowtie-shaped support is proportional to the diameter of the support of the object being imaged.  Thus, more compact objects can be adequately sampled with fewer projections.  Intuitively, the bowtie shape is a result of the limited slope of the lines that make up the sinogram.

When a digital signal is sampled, the Fourier series is periodic and one may assess the recoverable frequencies of the signal by observing the overlap of the translated copies of the original spectrum.  When the sampling is sufficient, the translated copies do not overlap.  Figure \ref{fig:fanBeamSampling} shows examples of sufficient sampling (middle) and the case where the angular sampling, $\beta$, is reduced by a factor of two (right).  We see here that when one under-samples in angle, the higher frequencies (in both angle and detector column) are corrupted, but the lower frequencies remain unaliased.  Thus, if the high frequency components are removed from the data, then the data can be adequately sampled at a lower rate.  This is the motivation behind our multi-step reconstruction process.

\subsection{Regularized Least Squares of Simple Functions} \label{sec:SimpleLS}

As previously described, a benefit of reconstructing low resolution volumes is the reduction of the number of unknowns.  A somewhat more sophisticated approach is to group voxels into sets, thus forming a simple function and forcing sets of voxels to share one (unknown) value.  For example, the unknown volume could be defined as a mesh on a point cloud.  The sets that make up the simple function can be defined by a segmentation of an initial reconstruction or by using a priori knowledge of the structure of the object being imaged.

The mathematical description of a simple function is given by
\begin{eqnarray*}
f_j &:=& \sum_k a_k \ind_{A_k}[j], \\
\ind_{A_k}[j] &:=& \begin{cases} 1, & \text{if } j \in A_i, \\ 0, & \text{otherwise}, \end{cases}
\end{eqnarray*}
where $\{a_k\}$ are the unknowns.

Then the RLS functional of this simple function can be written as
\begin{eqnarray}
\Phi(a) &:=& \left(Pf - g\right)^T\left(Pf-g\right) + \beta R(f) \notag \\
&=& \left(\sum_k a_kP\ind_{A_k} - g\right)^T\left(\sum_k a_kP\ind_{A_k} - g\right) + \beta R\left(\sum_k a_k \ind_{A_k}\right) \label{eq:SimpleLScost}
\end{eqnarray}
and the components of the gradient of this cost functional are given by
\begin{eqnarray*}
\frac{\partial}{\partial a_k}\Phi(a) &=& \sum_j \ind_{A_k}[j] \left[P^T(Pf-g) + \beta R'(f)\right]_j.
\end{eqnarray*}

If using a preconditioned conjugate gradient method to minimize this cost function, the step size can be calculated by
\begin{eqnarray*}
\lambda^n &:=& \frac{<u^n, d^n>}{<d^n, P^TPd^n + \beta R''(f^n)d^n>}.
\end{eqnarray*}

This method is a small modification of established techniques.  One advantage to this approach is that one can use a standard reconstruction voxel grid to define the image, and one can use the standard projection and backprojection algorithms with this data rather than writing specialized projection and backprojection routines as is the case for the point cloud methods.  The benefit of this method is that it can be used as a preliminary reconstruction of an object which can be refined by a standard reconstruction method in a second stage.

\subsection{Regularized Derivative Least Squares} \label{sec:RDLS}

The motivation for the method described in this section is that many CT images can be described as a low attenuation background with relatively sparse, high attenuation foreground.  The strategy is to first reconstruct the foreground of the CT image by leveraging its sparsity.  Since the foreground is sparse, few image voxels need to be reconstructed which leads to a low computational cost and fast reconstruction.  This should be followed by any reconstruction method seeded with the Regularized Derivative Least Squares (RDLS) reconstruction.

We make use of the RDLS functional which is given by
\begin{eqnarray}
\Phi_{RDLS}(f) &:=& \frac{1}{2}\left[\nabla(Pf - g)\right]^T\left[\nabla(Pf - g)\right] + \beta R(f) \notag \\
&=& \frac{1}{2}\left(Pf - g\right)^T\nabla^T\nabla\left(Pf - g\right) + \beta R(f) \notag \\
&=& -\frac{1}{2}\left(Pf - g\right)^T\Delta\left(Pf - g\right) + \beta R(f) \label{eq:RDLScost}
\end{eqnarray}
where we have used $\nabla^T\nabla = -\Delta$. The gradient and Hessian of this cost functional is given by
\begin{eqnarray*}
\Phi_{RDLS}'(f) &=& -P^T\Delta (Pf-g) + \beta R'(f) \\
\Phi_{RDLS}''(f) &=& -P^T \Delta P + \beta R''(f).
\end{eqnarray*}
Here $\nabla$ can be a one or two-dimensional gradient on each projection (derivative is not taken over angle).  If $R$ is convex, then this functional is convex because its Hessian is positive semi-definite
\begin{eqnarray*}
<-P^T\Delta P f, f> = <-\Delta Pf, Pf> = <\nabla Pf, \nabla Pf> = \| \nabla Pf \|^2 \geq 0.
\end{eqnarray*}

Suppose that $f_{RDLS}$ is an RDLS solution.  Then the high attenuation objects are mostly removed by the difference $g - Pf_{RDLS}$ and the transition into this modified region is smooth.  It is smooth because the edges are suppressed by the design of the cost function.  Thus when one seeds a full frame iterative reconstruction algorithm with $g - Pf_{RDLS}$ there are much fewer attenuation streak-generating gradients, or the magnitude of these gradients is diminished.  The reduction of these streak artifacts is the direct motivation for this approach.

Note that $-\Phi_{RDLS}'(0) = -P^T \Delta g$ which is roughly equal to a Lambda or local tomography reconstruction \cite{Faridani_SIAM_92a,Faridani_SIAM_97,Katsevich_SIAM_99}.

\subsubsection{Relationship to Least Squares}

Suppose we have parallel beam CT data, and we are in the continuous domain.  Then we have that $P^T\Delta = \Delta P^T$, where the second $\Delta$ is over image (Euclidean) space.  Let the Least Squares and Derivative Least Squares functionals be denoted by
\begin{eqnarray*}
\Phi_{LS}(f) &:=& \frac{1}{2}\left(Pf - g\right)^T\left(Pf - g\right) \\
\Phi_{DLS}(f) &:=& \frac{1}{2}\left[\nabla(Pf - g)\right]^T\left[\nabla(Pf - g)\right].
\end{eqnarray*}
Also let $f_{LS}$ be a least squares solution.  Then $P^Tg = P^TPf_{LS}$, so $\Delta P^Tg = \Delta P^TPf_{LS}$ and thus $P^T \Delta g = P^T \Delta Pf_{LS}$.  Therefore $\Phi_{DLS}'(f_{LS}) = 0$ and $f_{LS}$ is a derivative least squares solution.  It is not necessarily true that a derivative least squares solution is a least squares solution.  Similarly, one can show that
\begin{eqnarray*}
<\Phi_{LS}'(f), \Phi_{DLS}'(f)> \; \geq \; 0.
\end{eqnarray*}
Thus we see that there is much similarity between the least squares and derivative least squares solution.  We also see that the gradient descent algorithm applied to either cost functional will reduce the cost for either.

\subsection{Exploiting Histogram Sparsity} \label{sec:HistogramSparsity}

Industrial objects are often composed of a small set of materials and the density of each material has only minor density changes.  Thus the histogram of such object's LAC distribution is sparse: it should be composed of a small collection of tight distributions around each material in the object.  When the object is composed of only one material, one may consider the reconstruction problem to be a discrete tomography problem \cite{Gabor_DT_2012, Gardner_DT_1997}.  A common discrete tomography algorithm is the Discrete Algebraic Reconstruction Technique (DART)\cite{DART}.

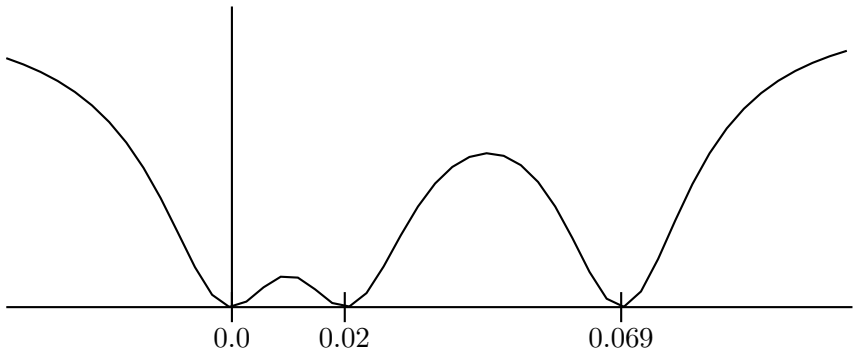
\begin{figure}[h!]
\psset{xunit=75cm,yunit=4cm}
\begin{pspicture}(-0.07,-0.1)(0.04,1.2)
\psplot{-0.04}{0.109}{x x mul x x mul 0.0002 add div x 0.02 sub 2 exp x 0.02 sub 2 exp 0.0002 add div mul x 0.069 sub 2 exp x 0.069 sub 2 exp 0.0002 add div mul}
\psline(-0.04,0.0)(0.11,0.0) \psline(0.0,0.0)(0.0,1.0) 
\rput(0,-0.1){0.0} \rput(0.02,-0.1){0.02}
\psline(0,-0.05)(0,0.05) \psline(0.02,-0.05)(0.02,0.05)
\psline(0.069,-0.05)(0.069,0.05) \rput(0.069,-0.1){0.069}
\end{pspicture}
\caption{An example of a histogram sparsity loss function with target values $\mu = (0.0, 0.02, 0.069)$.} \label{fig:histogramSparsitiyLossFunction}
\end{figure}

We propose a new method that uses a soft constraint to encourage histogram sparsity.  This method introduces a new additive term to a cost function given by
\begin{eqnarray}
\Psi(f; \mu) &:=& \alpha \| u_\mu(f) \|_1,
\end{eqnarray}
where $\mu = (\mu_1, \dots, \mu_n)$ are the expected LAC values of the materials in the object and $u_\mu$ is a non-negative, smooth function that is only equal to zero at $\{\mu_i\}_{i=1}^n$ and smoothly approaches a positive value away from $\{\mu_i\}_{i=1}^n$.  For example, one form for this loss function is given by
\begin{eqnarray*}
u_\mu &=& \prod_k \frac{(f - \mu_k)^2}{(f - \mu_k)^2 + \frac{1}{2}m^2}, \\
m &:=& \text{min}_k \left|\mu_k - \mu_{k+1}\right|.
\end{eqnarray*}
A plot of this loss function is shown in Figure \ref{fig:histogramSparsitiyLossFunction}.  One may treat $\{\mu_i\}_{i=1}^n$ as fixed parameters or as unknowns that are optimized with $f$.  This functional encourages voxel values to approach one of the target $\mu_i$ values which increases the histogram sparsity in the reconstruction.

\subsection{Exploiting Azimuthal Sparsity} \label{sec:AzimuthalSparsity}

Many industrial objects exhibit symmetry, especially azimuthal (cylindrical) symmetry.  Flaws in the manufacturing process (voids, cracks, or warping) may introduce slight asymmetries.  For these objects we construct a TV-like loss function that attempts to sparsify the gradient in the azimuthal direction.  Thus let $v_\varphi$ be a low pass filter in the azimuthal direction and let $h$ be any loss function.  Then one may construction an azimuthal sparsity penalty by
\begin{eqnarray}
\Psi(f; \varphi) &:=& \alpha \| h(v_\varphi(f) - f) \|_1. \label{eq:azimuthalSparsityFunctional}
\end{eqnarray}
We have found that a moving average filter works well for $h_\varphi$.  As discussed in Section \ref{sec:samplingTheory}, the strength of the aliasing from undersampling increases with distance from the center of rotation.  The strength of this azimuthal penalty term increases with distance from the center of rotation and thus is better adaptive to under-sampling artifacts that traditional regularization approaches that are spatially-invariant.

Azimuthal blurring has found other applications such as attenuation correction of PET data with sparse-view CT \cite{DeMan_ULD_PMB_2015}.  Note that our azimuthal sparisty method does not require the object to be centered in the field of view of the CT data as do view interpolation algorithms.  One does need to specify the center of the object in the reconstruction space, but this can almost always be estimated very accurately.

\subsection{Estimation of the Image Foreground without Performing a Reconstruction}

The Simple Function RLS and some uses of RDLS approaches to few-view reconstruction require that the voxels composing the image foreground be identified.  If there are enough projections and the scanner geometry is such that one can reconstruct the image with Filtered Backprojection, then one can find the image foreground by various image segmentation techniques.  Otherwise, one must have a method to efficiently (quickly) identify the image foreground.  In this section we describe a method to estimate the foreground regions of the CT image without performing a reconstruction.

To form a map of the image foreground, first segment out the highly attenuated rays of the projection data and form a binary foreground mask denoted by $g_{FG}$, where $g_{FG}$ is equal to one for the highly attenuating rays and zero otherwise.  Then a binary foreground mask in image space can be found by
\begin{eqnarray}
f_{FG} &:=& 1 - u\left( P^T(1 - g_{FG}) \right), \label{eq:spaceCarving} \\
u(t) &:=& \begin{cases} 1, & t > 0, \\ 0, & \text{otherwise}. \end{cases} \notag
\end{eqnarray}
We refer to this as a space-carving algorithm because the backprojection process carves out the space that is identified in the projection mask, $g_{FG}$.  Note that simply back projecting the foreground mask, $g_{FG}$, will result in streaks and will not compactly identify the image-space foreground.  One can then split this foreground mask, $f_{FG}$ into different regions by a connected component algorithm for use by the Simple Function RLS algorithm.

\section{Numerical Experiments}

In this section we illustrate uses for the Simple Function RLS, RDLS, histogram sparsity prior, and azimuthal regularization few-view reconstruction methods.  All numerical experiments were performed using the Livermore Tomography Tools (LTT) software package \cite{LTTpaper}.  These algorithms can also be found in the LEAP-CT \cite{LEAPpaper} open source library as well.  Our simulated data is generated by ray-tracing through geometric solids, not projections through a voxelized volume.  This is an important distinction because it is misleading to demonstrate reconstruction algorithms on data generated using the same forward model used in the reconstruction method.  When possible, we demonstrate our methods with measured data to illustrate their effectiveness in practice.  Many algorithms appearing in the literature may work well with simulated data, but fail when applied to real data.

\begin{figure}[h!]
\begin{center}
\begin{tabular}{ccc}
FBP & ASD-POCS & Simple Function RLS \\
\includegraphics[scale=0.15]{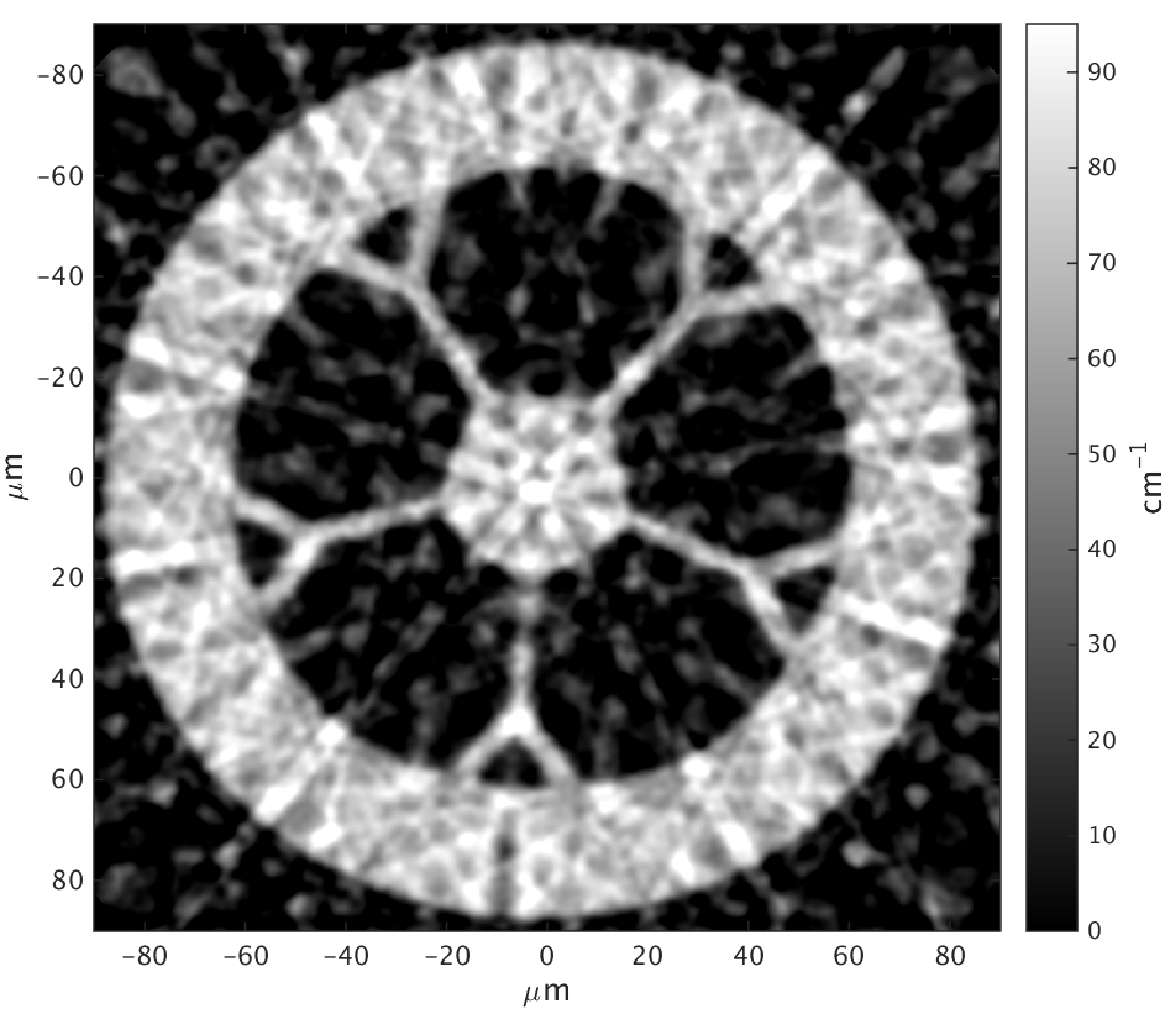}
& \includegraphics[scale=0.15]{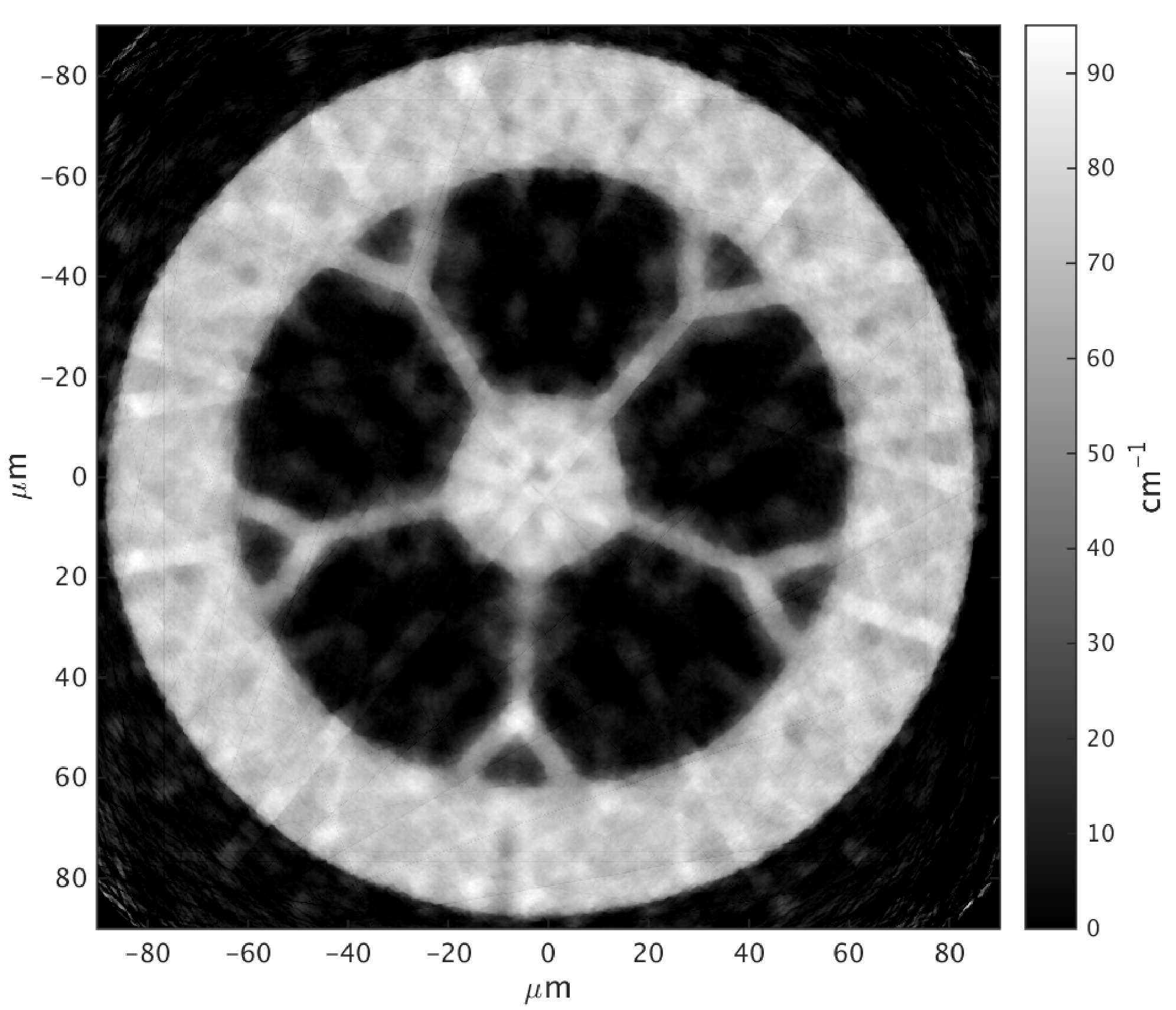}
& \includegraphics[scale=0.15]{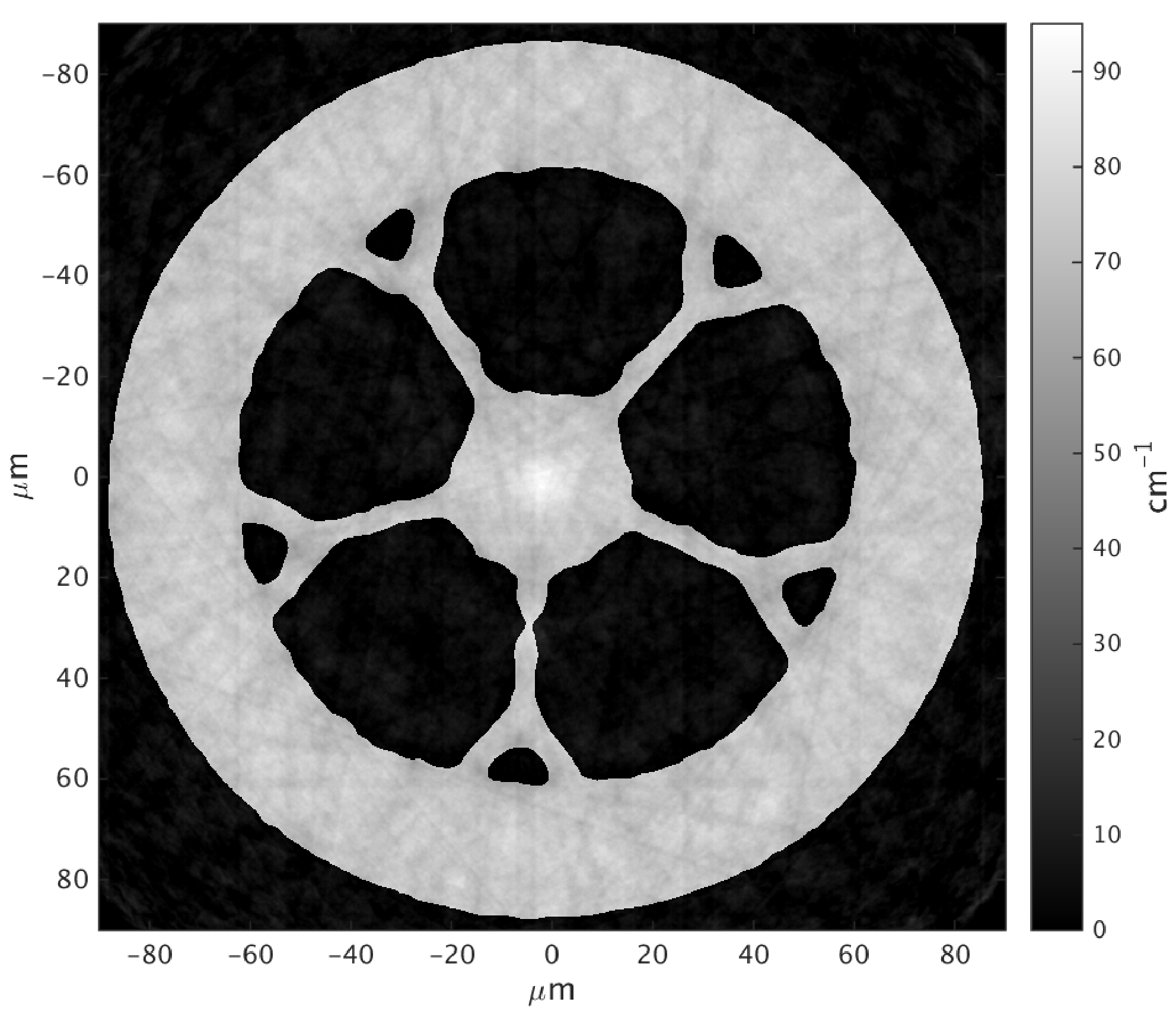}
\end{tabular}
\caption{Reconstructed slices of 19 measured CT projection of a 180 um diameter glass fiber.  Notice that the Simple Function RLS image is the only image where the germanium-doped core at the center of the image is visible.} \label{fig:simpleFunctionLS}
\end{center}
\end{figure}

In our first example, we imaged a 180 um diameter glass fiber with 19 projections on an Xradia Ultra XRM scanner at LLNL.  The goal was to recover the average density and the density variations within the fiber.  Our goal was efficient recovery of the average density and the density variations within the fiber, as inspection was required for hundreds of fibers.  To reduce scan time, we only collected 19 projections for each fiber.  The data was first reconstructed with FBP using a smooth ramp filter.  The region occupied by the glass fiber was segmented using an active contour segmentation algorithm \cite{ChanVese_2001}.  Next the data was reconstructed with the Simple Function RLS algorithm, where all voxels within the identified glass fiber region were grouped in the same set and all remaining voxels were independent.  The final volume was reconstructed using the standard least squares method, but iterations were seeded with the Simple Function RLS volume.  We also reconstructed the data with ASD-POCS which is a state-of-the-art few-view CT volume reconstruction algorithm for reference.  The results are shown in Figure \ref{fig:simpleFunctionLS}.  Notice that the Simple Function RLS image is the only image where the germanium-doped core at the center of the image is visible.

\begin{figure}[h!]
\begin{center}
\begin{tabular}{cc}
Bullet Penetrating Aluminum Plate & Explosive Charge (early time) \\
\includegraphics[width=0.45\textwidth]{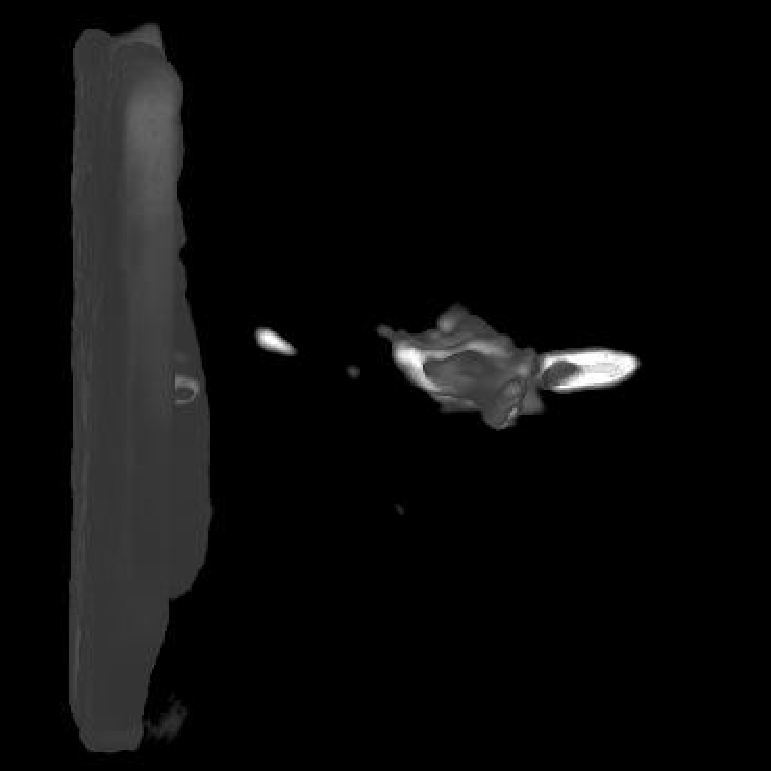}
& \includegraphics[width=0.45\textwidth]{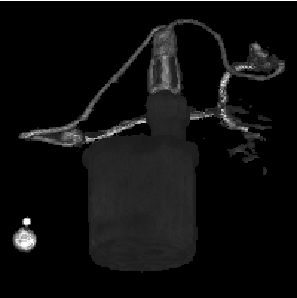} \\
Explosive Charge (early time) & Explosive Charge (late time) \\
\includegraphics[width=0.45\textwidth]{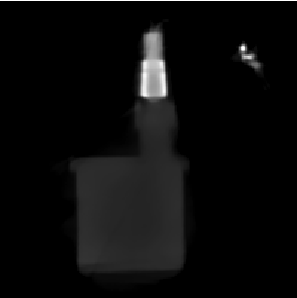}
& \includegraphics[width=0.45\textwidth]{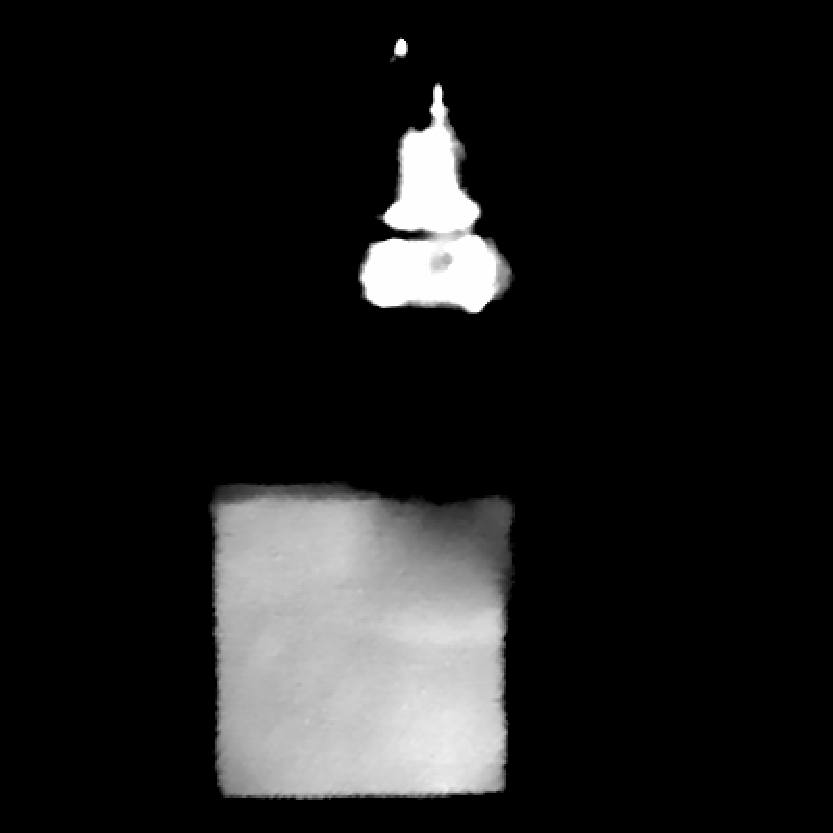}
\end{tabular}
\end{center}
\caption{Reconstruction of flash x-ray CT data at DEVCOM ARL using the RDLS algorithm.  (top left) 3D rendering of a bullet passing through an aluminum plate, where it is possible to observe the bullet jacket has been stripped from its core.  (top right) 3D rendering of an explosive charged detonator at an early time when the detonation front is still in the booster.  (bottom left) Same as top right, but showing a reconstructed slice.  (bottom right) Same as bottom left, but here the imaging was taken at a later time when the detonation front was in the main charge.} \label{fig:ARLreconstructions}
\end{figure}

We demonstrate our RDLS algorithm on data collected on the Multi-Energy Flash CT (MEFCT) system at the DEVCOM Army Research Laboratory (DEVCOM ARL) \cite{Zellner_MEFCT_2018}.  This system consists of five 150 kV source-detector pairs, five 300 kV source-detector pairs, and five 450 kV source-detector pairs.  Reconstructions of dynamic events imaged on this system are shown in Figure \ref{fig:ARLreconstructions}.  In these data sets, most projections are truncated, i.e., some part of the object falls outside the projection.  Due to the data truncation and the lack of quantitative accuracy associated with the shot-to-shot variances of flash radiography, the RDLS algorithm was the only algorithm capable of reconstructing these datasets.  For example, applying ADS-POCS diverged on the second iteration.

\begin{figure}[h]
\begin{center}
\begin{tabular}{cc}
ASD-POCS & RDLS + ASD-POCS \\
\includegraphics[scale=0.35]{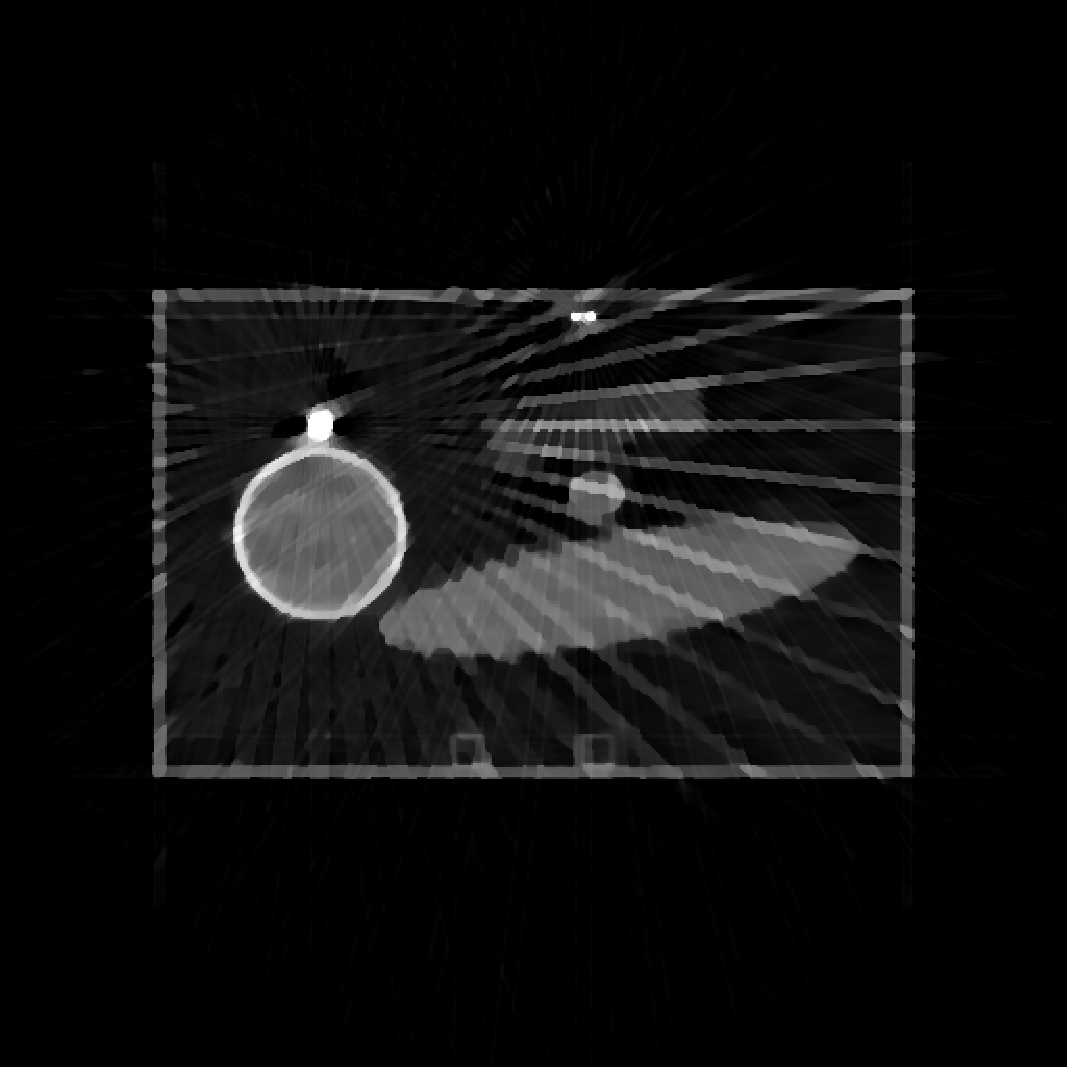}
& \includegraphics[scale=0.35]{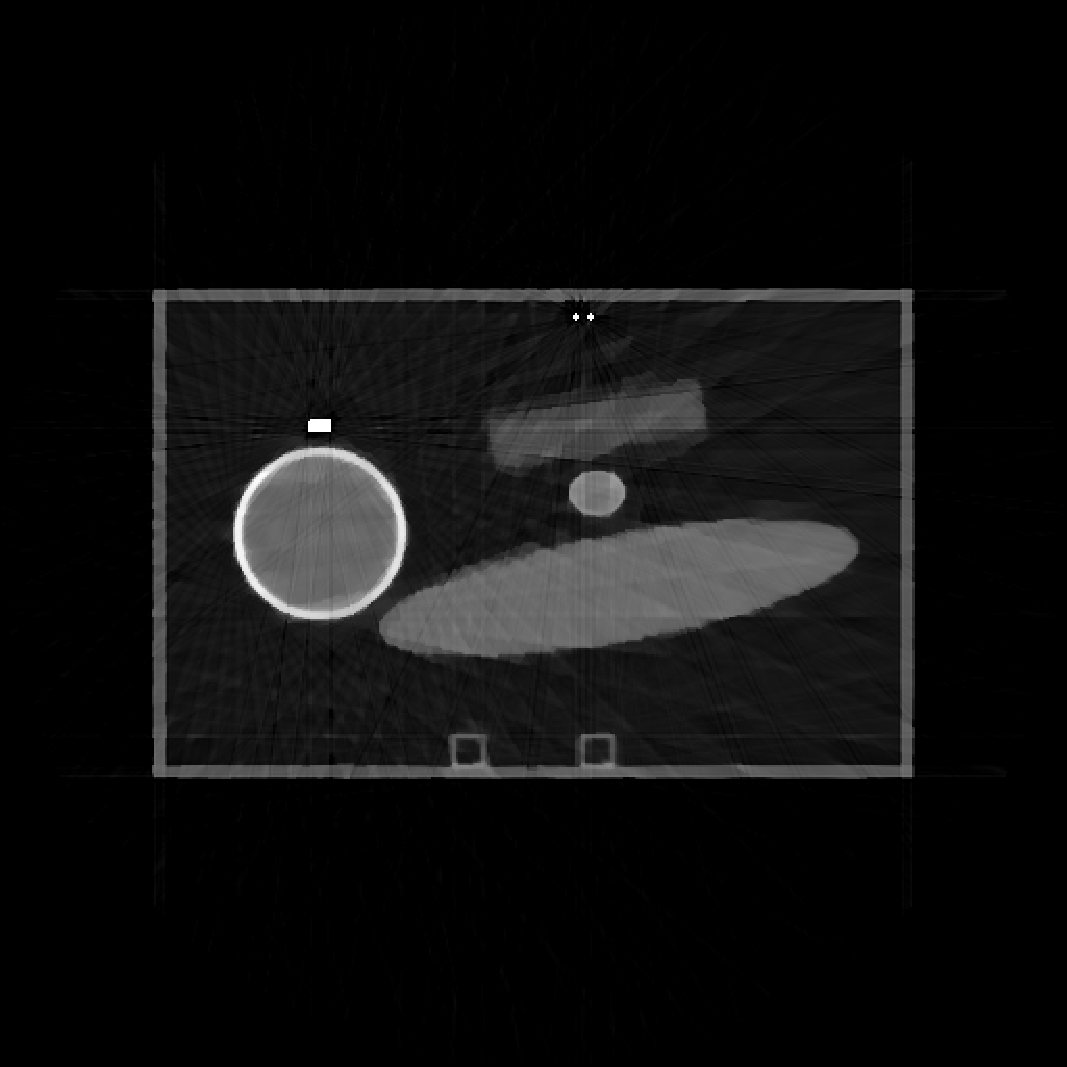}
\end{tabular}
\caption{Reconstructed slices from 28 simulated projections.} \label{fig:RDLS}
\end{center}
\end{figure}

For our first RDLS simulation test, we simulated 28 fan beam projections.  The phantom includes a water bottle in glass, a piece of rubber, a stack of paper, a small iron box, and two copper wires.  The voxels in the image that contain the small iron box and copper wires were identified by the space carving algorithm (equation \ref{eq:spaceCarving}).  Then we performed a RDLS reconstruction on the iron box and copper wire voxels.  Finally, we reconstructed the data with ASD-POCS and seeded the iterations with the RDLS image.  The results of simply reconstructing the data with the ASD-POCS algorithm and with the combination of the RDLS and ASD-POCS algorithms are shown in Figure \ref{fig:RDLS}.  Streaks in the image produced by RDLS + ASD-POCS are not as pronounced as the streaks in the image reconstructed only with the ASD-POCS algorithm.

\begin{figure}[h]
\begin{center}
\begin{tabular}{cc}
ASD-POCS & RDLS+ASD-POCS \\
\includegraphics[scale=0.35]{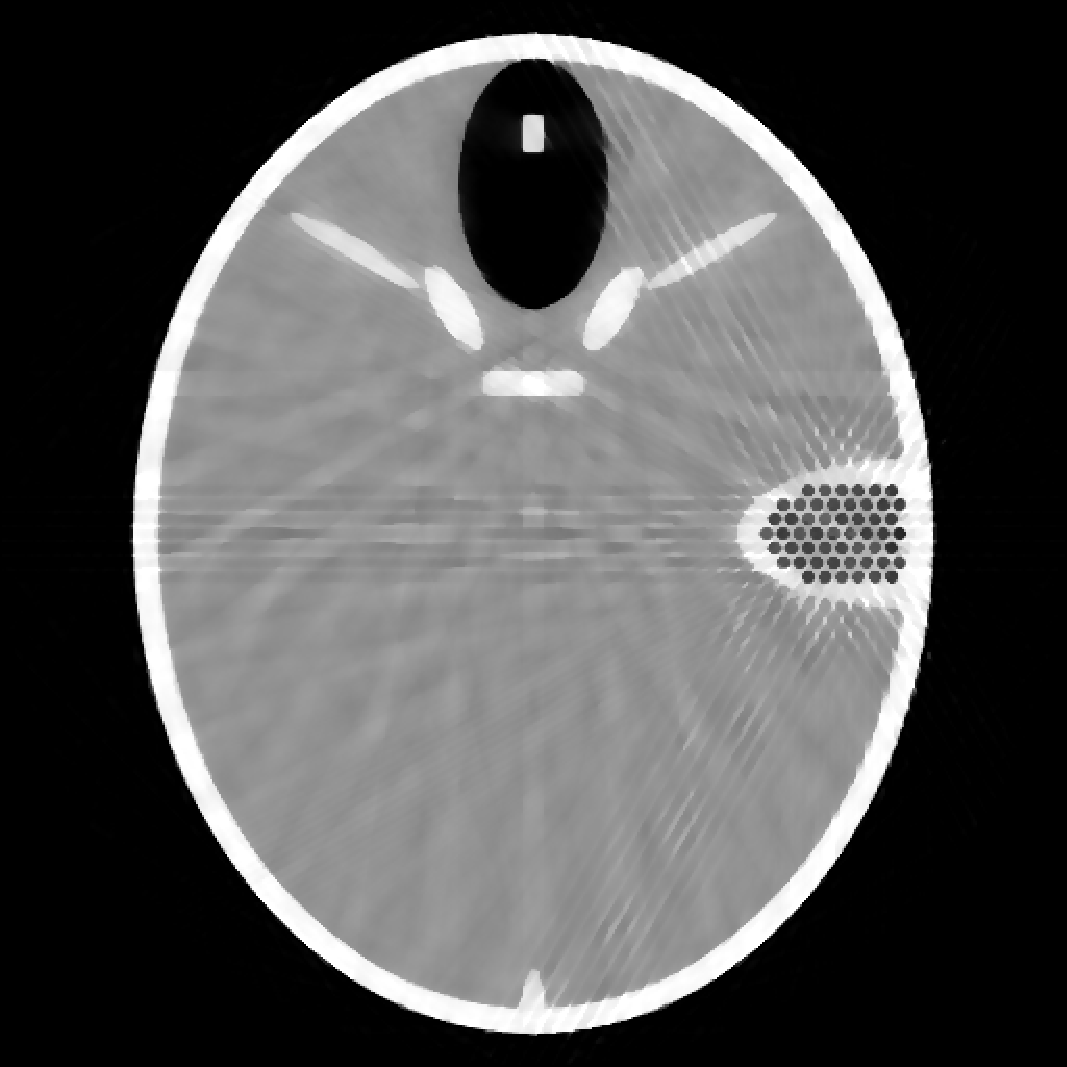}
& \includegraphics[scale=0.35]{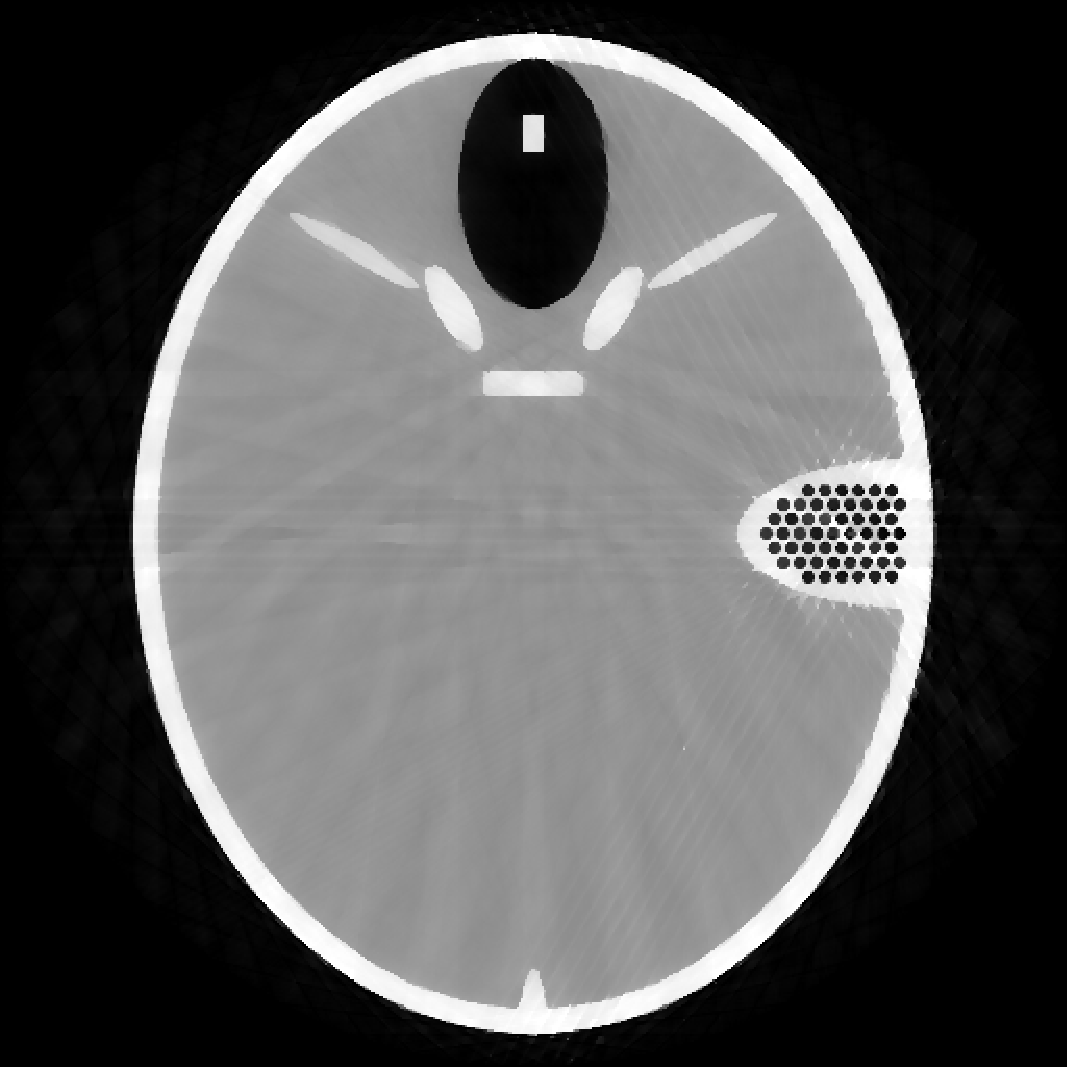} \\
\includegraphics[scale=1.4]{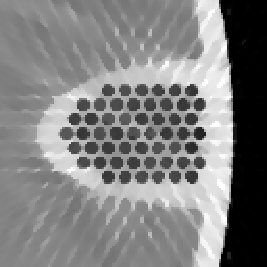}
& \includegraphics[scale=1.4]{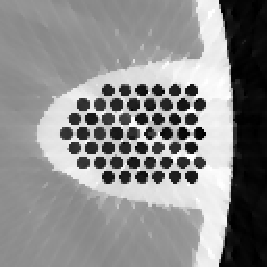}
\end{tabular}
\caption{Reconstructed slices from 24 simulated fan-beam projections.} \label{fig:FORBILD}
\end{center}
\end{figure}

In a second RDLS simulation test, we simulated 24 fan beam projections of the FORBILD head phantom.  First, we performed an ASD-POCS reconstruction.  Our second reconstruction started by reconstructing the ear region only with RDLS and then reconstructing the whole object with ASD-POCS.  Results are shown in Figure \ref{fig:FORBILD}.  Notice that the RDLS+ASD-POCS reconstruction has fewer streaks and better contrast than the ASD-POCS reconstruction.  The reduced prominence of the streaks was a result of better recovery of the edges, especially in the ear region.  Note that the RWLS algorithm penalizes a weighted squared difference of the forward projection of the reconstruction and the measured data, while the RDLS algorithm penalizes a weighted squared difference of the gradient of the forward projection of the reconstruction and the measured data.  By using a gradient in the formulation, the RDLS algorithm does a better job in estimating transition regions.  

The RDLS reconstruction may be used as a stand-alone reconstruction algorithm, but low frequency image features converge very slowly.  Note that this is exactly the opposite of most CT iterative reconstruction algorithms such as RWLS, SART, or ASD-POCS.  To improve convergence of the low-frequency components, one can either use another iterative reconstruction before or after RDLS or one may use a low-pass filter as a preconditioner for RDLS.  A simple Gaussian filtered works well.

Next we demonstrate the effectiveness of our histogram sparsity prior on a CT dataset of a foam sample collected at the Advanced Light Source (ALS) synchrotron facility of Lawrence Berkley National Laboratory.  Here we use only 15 of the 1440 projections to perform reconstructions using the ASD-POCS and RWLS with histogram sparsity prior.  We note that the ASD-POCS reconstruction has significant artifacts, but the reconstruction utilizing the histogram sparsity is much closer to the reconstructing using all 1440 projections.

\begin{figure}[h!]
\begin{center}
\begin{tabular}{ccc}
FBP 1440 views & ASD-POCS 15 views & histogram sparsity 15 views \\
\includegraphics[width=0.3\textwidth]{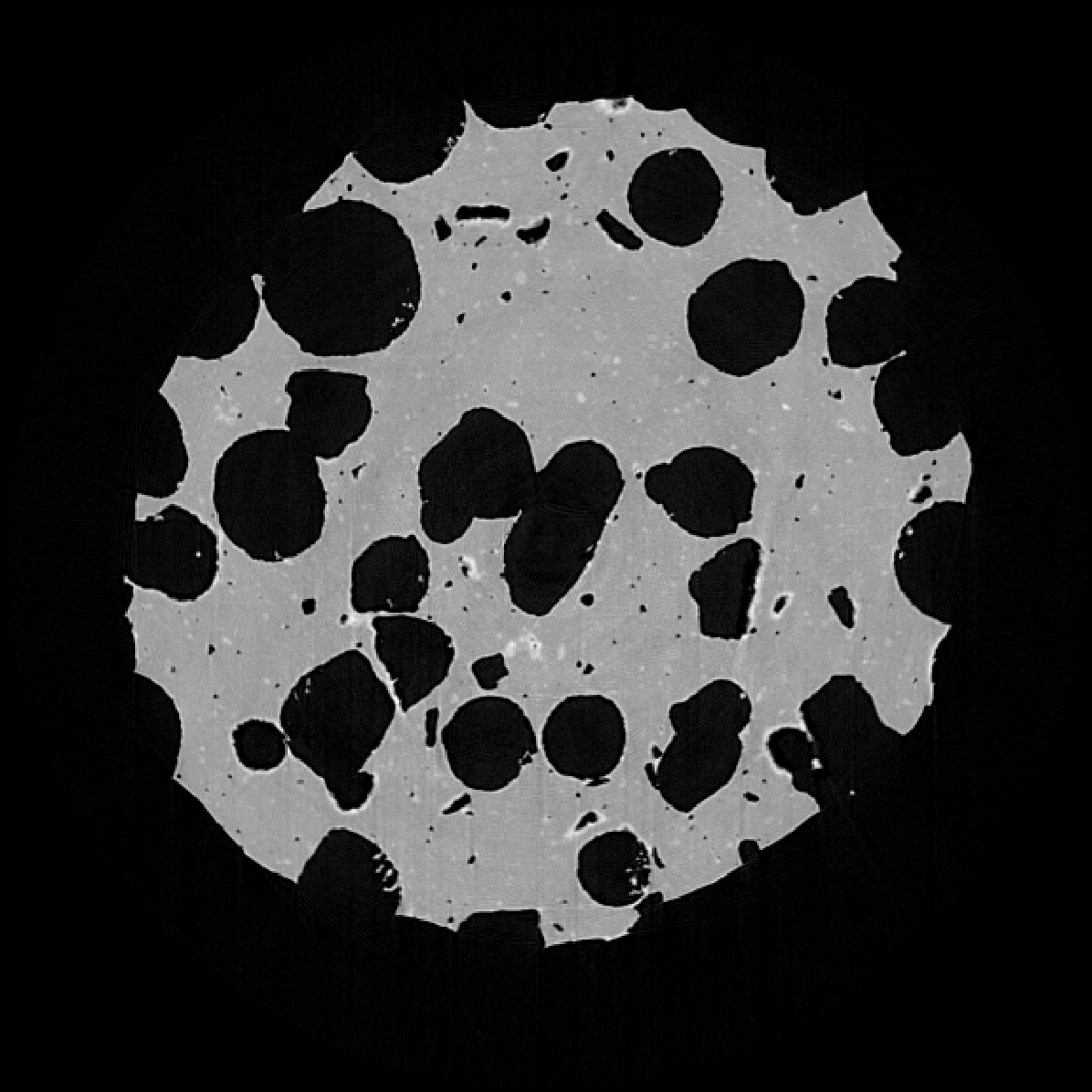}
& \includegraphics[width=0.3\textwidth]{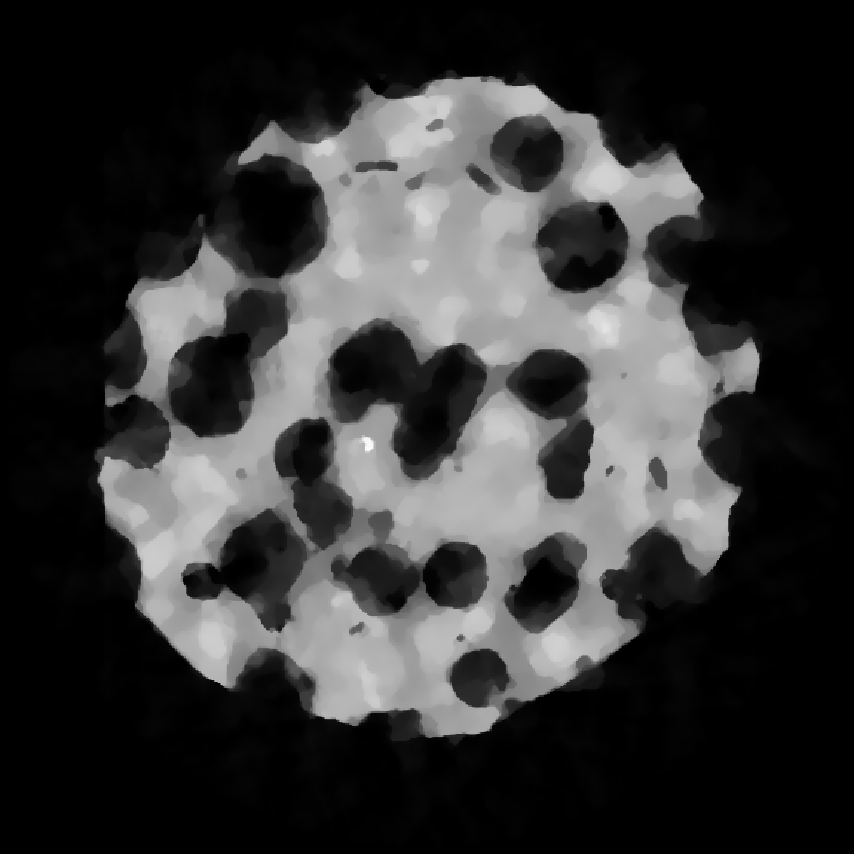}
& \includegraphics[width=0.3\textwidth]{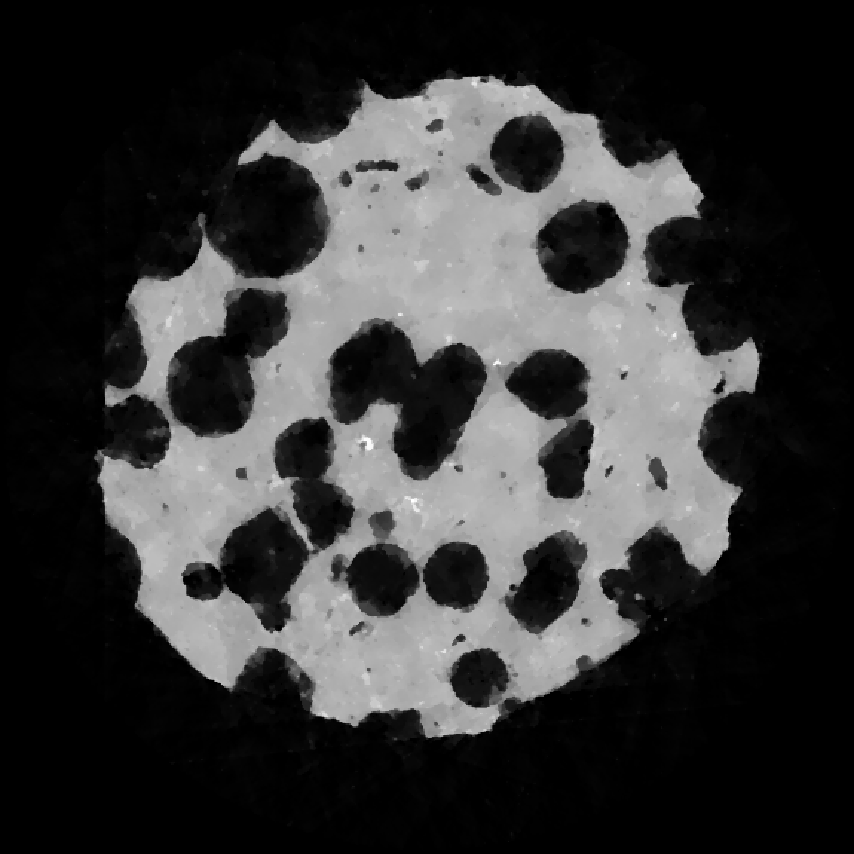}
\end{tabular}
\end{center}
\caption{Reconstruction of CT data of a foam sample from ALS.} \label{fig:ALSfoam}
\end{figure}

\begin{figure}[h!]
\begin{center}
\begin{tabular}{cc}
FBP & FBP with azimuthal blur post-processing \\
\includegraphics[width=0.5\textwidth]{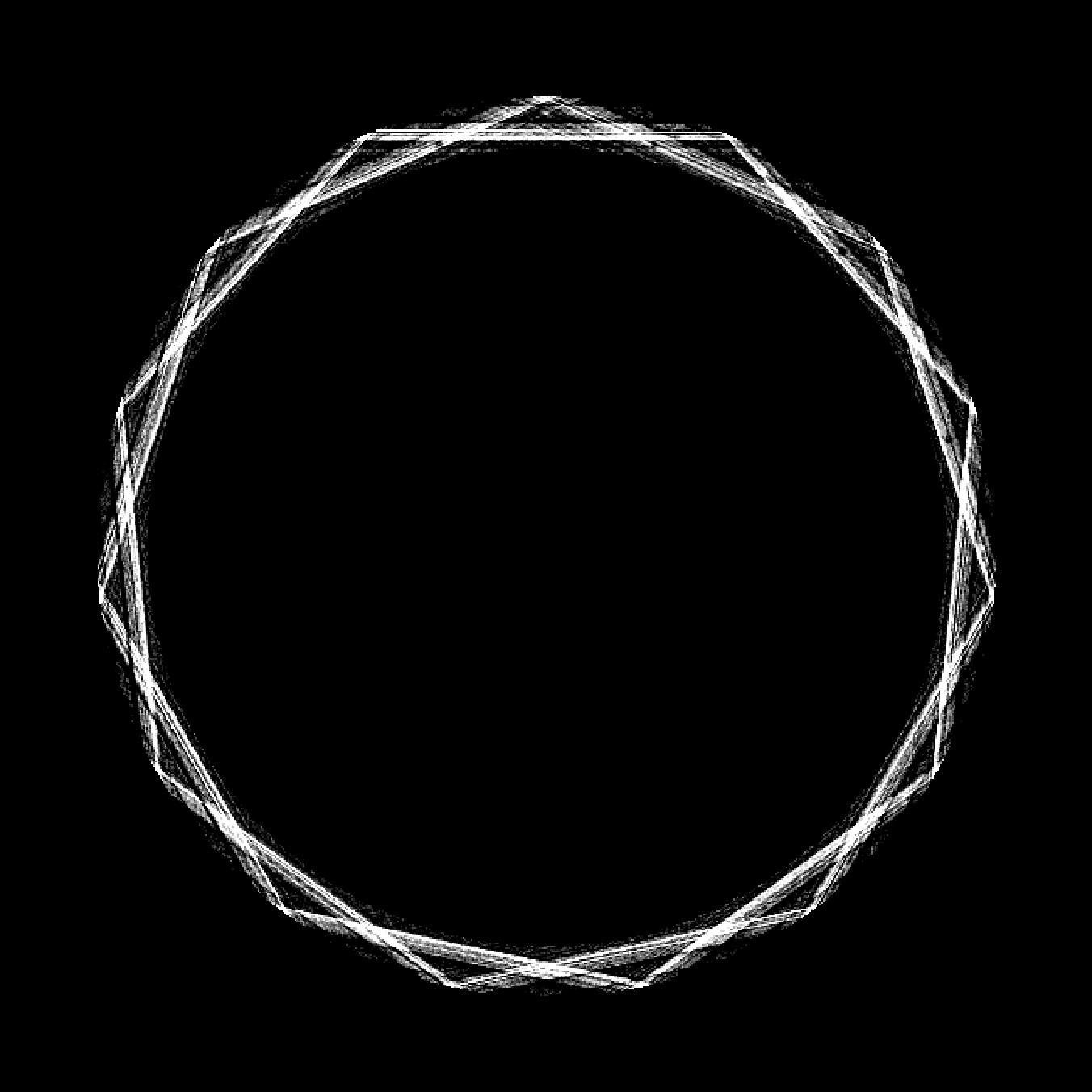}
& \includegraphics[width=0.5\textwidth]{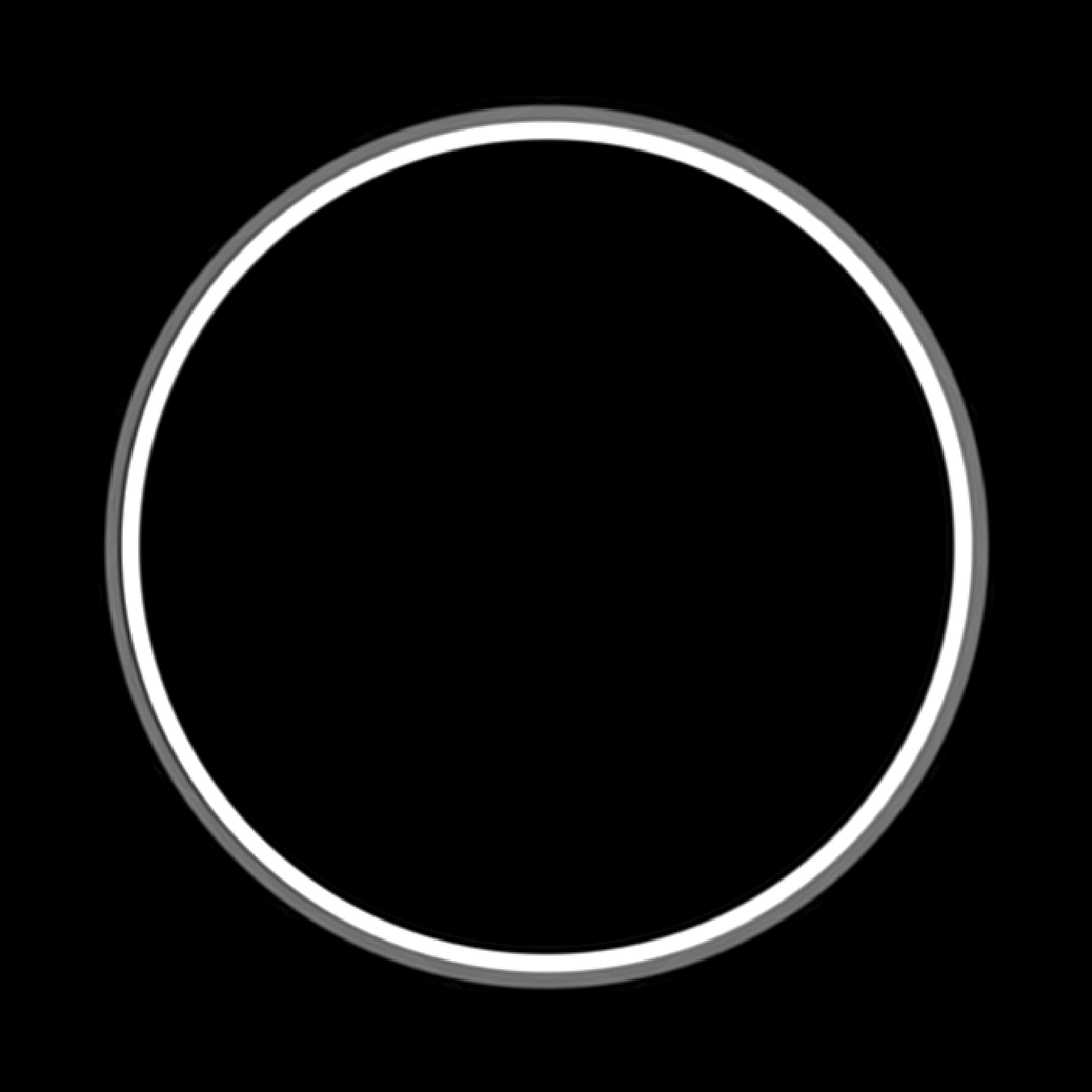} \\
RLS & RLS with azimuthal sparsity \\
\includegraphics[width=0.5\textwidth]{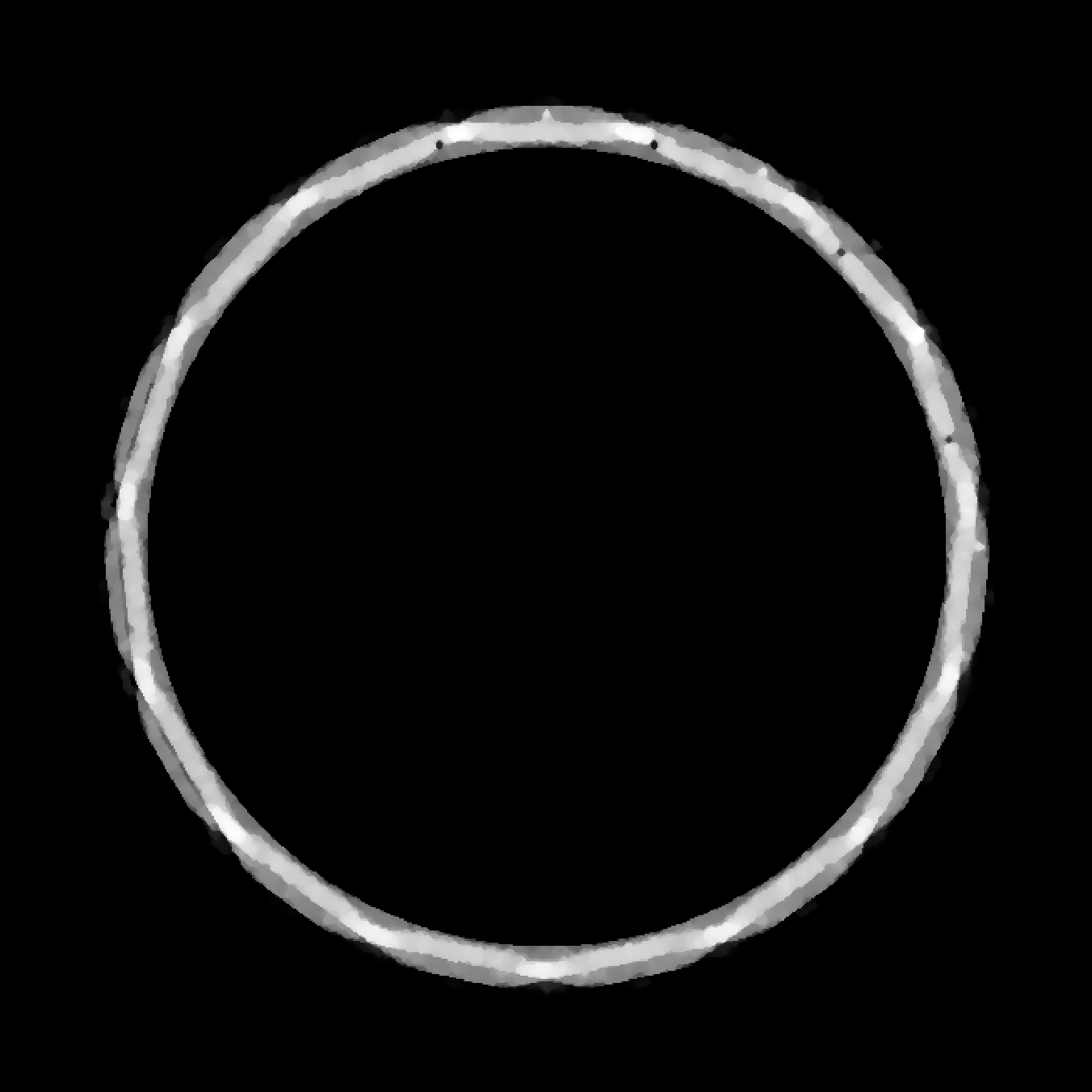}
& \includegraphics[width=0.5\textwidth]{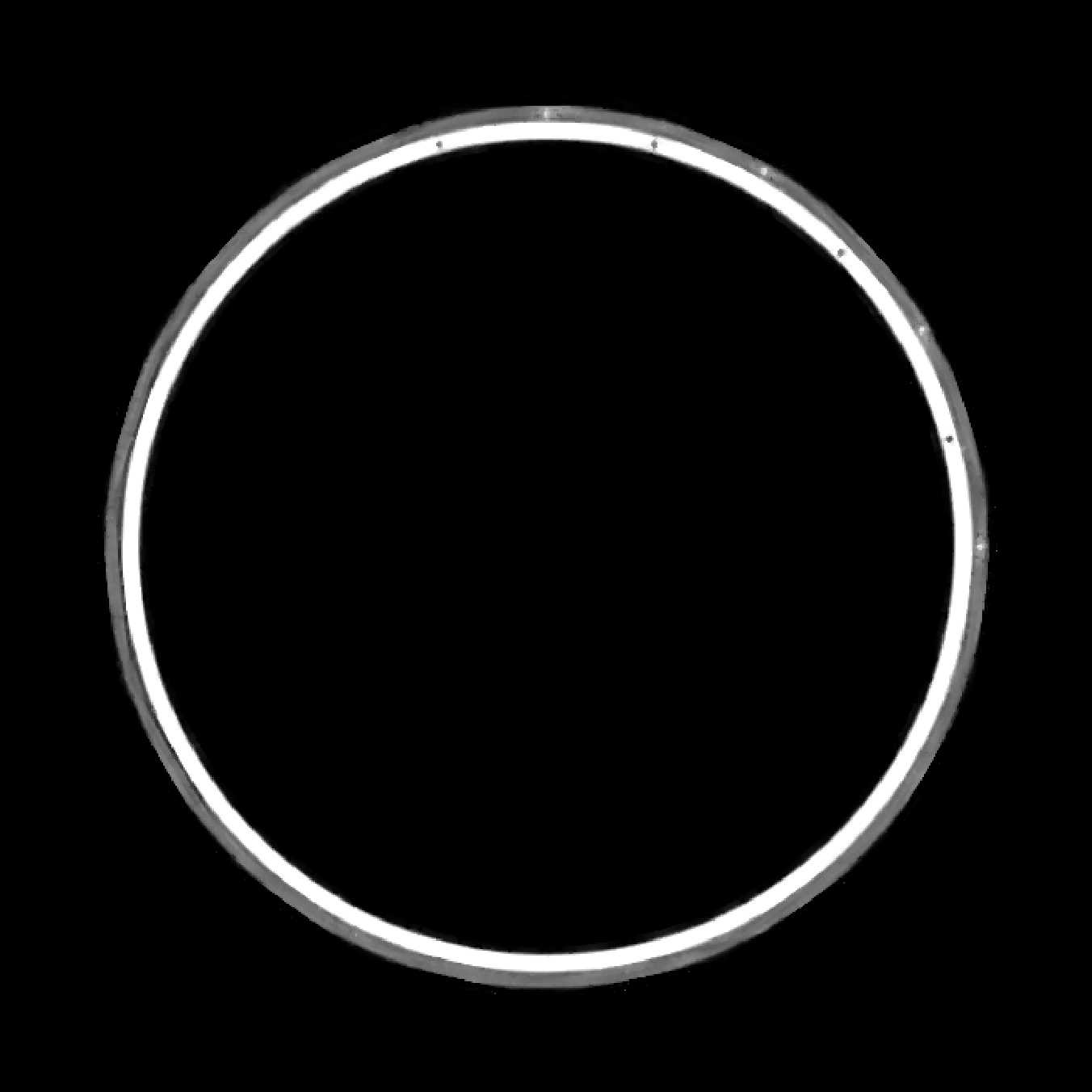}
\end{tabular}
\end{center}
\caption{Reconstructions of a 15-view cone-beam simulated dataset of a two-layer pipe phantom.} \label{fig:pipe}
\end{figure}

Our final demonstration is for the effectiveness of the azimuthal sparsity regularization.  We simulated 15 noisy cone-beam projections of a two-layer pipe phantom.  There is an air gap between the two layers on the left side of the object that spans 70$^\circ$ around the pipe.  There are also voids on the inner layer and high density inclusions on the outer layer.  We reconstructed the data in four different ways (1) FBP, (2) post-processed the FBP reconstruction by applying a 45 degree azimuthal blur, $v_\varphi(f)$, in equation (\ref{eq:azimuthalSparsityFunctional}), (3) RLS with anisotropic TV regularization, and (4) RWLS with anisotropic TV regularization and the azimuthal sparsity (blur kernel is 4$^\circ$ in size) regularization.  Results are shown in Figure \ref{fig:pipe}.  The FBP reconstruction with azimuthal blur post-processing recovered the gap, but not the voids or inclusions.  The RLS reconstruction had significant artifacts and did not recover the gap.  The full 70$^\circ$ gap and all voids and inclusions are recovered in the RLS reconstruction with azimuthal sparsity regularization.

\section{Conclusion and Discussion}

This paper describes several efficacious methods for few-view CT reconstruction.  Algorithms were demonstrated with measured data to illustrate their effectiveness in real imaging problems.  These methods demonstrate significant performance improvements over other state of the art techniques for few-view CT reconstruction.  The novel algorithms developed here are
\begin{enumerate}
\item Simple Function Least Squares reconstruction algorithm (Section \ref{sec:SimpleLS})
\item Regularized Derivative Least Squares (Section \ref{sec:RDLS})
\item Histogram Sparsity Regularization (Section \ref{sec:HistogramSparsity})
\item Azimuthal Sparsity Regularization (Section \ref{sec:AzimuthalSparsity})
\end{enumerate}
Each of these methods leverages a set of assumptions on the object being imaged to mitigate streak artifacts and produce quantitatively accurate reconstructions.  All these methods are available in the LEAP-CT \cite{LEAPpaper} open source library.

The RDLS algorithm has applications beyond few-view CT image reconstruction.  The semi-local nature of the RDLS cost function allows concentration of effort on subregions, while classic approaches to CT reconstruction are global by their nature.  RDLS thus permits piecewise refinement of the reconstructed image in selected subregions.  This approach can be useful in scenarios where CT projections are truncated, i.e., in so-called interior tomography \cite{Faridani_MSRI_03,Kudo_PMB_2008,Wang_TMI_2011} conditions.  The approach may be necessary, for example, to improve image quality in a region of interest or to reduce partial volume \cite{Pelc_MedPhys_1980,DeMan_TNS_2000} and metal streak artifacts \cite{Pelc_MedPhys_1981,Kalender_Radiology_1987,Zhao_TMI_2000,Bal_MedicalPhysics_2006,Pelc_SPIE_2011,Kachelriess_MedPhys_2012,Fleischmann_Radiology_2011,Seco_PhysMedBiol_2012,Koehler_CTmeeting_2011,Oehler_MIC_2006,Nuyts_TMI_2009,DeMan_TMI_2001,Karimi_ICASSP_2014}.  Also since fewer voxels are affected, the algorithm may be extremely computationally efficient.

\section{Acknowledgment}

This work was supported in part by the U.S. Science and Technology Directorate of the Department of Homeland Security (DHS) and in part by the Department of Energy under Contract HSHQDC-14-X-B0001.  This work was also supported by the Laboratory Directed Research and Development Program under LLNL-LDRD 18-ERD-052.  This work was partially funded under the DEVCOM ARL mission funds AH-80.


\bibliographystyle{unsrt}
\bibliography{bibliographyAll}

\end{document}